\documentclass[a4,11pt]{article}

\usepackage{blindtext}
\usepackage[a4paper, total={16.5cm, 23.5cm}]{geometry}

\RequirePackage{amsmath}
\RequirePackage{graphicx}

\newcommand{\lambdabar}{{\mkern0.75mu\mathchar '26\mkern -9.75mu\lambda}}
\RequirePackage{xcolor}

\normalsize
\title{ \bf LUXE: A new experiment to study non-perturbative QED  \\ in electron-laser and photon-laser collisions}

\author {Aharon Levy \\ Tel Aviv University,
  Tel Aviv-Yafo 6997801, Israel \\ \\
on behalf of the LUXE Collaboration}

\date{}

\begin{document}
\maketitle

\abstract{
The LUXE experiment (Laser Und XFEL Experiment) is an experiment in planning at DESY Hamburg using the electron beam of the European XFEL (Eu.XFEL). LUXE is intended to study collisions between a high-intensity optical laser pulse and 16.5 GeV electrons from the Eu.XFEL electron beam, as well as collisions between the laser pulse and high-energy secondary photons. This will elucidate Quantum Electrodynamics (QED) at the strong-field frontier, where the electromagnetic field of the laser in the electron rest frame is above the Schwinger limit. In this regime, QED is non-perturbative in the charge field coupling. This manifests itself in the creation of physical electron-positron pairs from the QED vacuum, similar to Hawking radiation from black holes. LUXE intends to measure the positron production rate in an unprecedented laser intensity regime. It is expected to start running in 2025. An overview of the LUXE experimental setup and its challenges and progress will be given, along with a discussion of the expected physics reach in the context of testing QED in the non-perturbative regime.
}

\vspace*{4cm}
   {\it Presented at DIS2022: XXIX International Workshop on Deep-Inelastic Scattering and Related Subjects, Santiago de Compostela, Spain, 
  2 - 6 May, 2022}

\newpage
\section{Introduction}
QED is the most precisely tested theory in the Standard Model of constituents of matter and their interactions. Its precise predictions  were made possible by calculations with perturbative methods due to the small value of the fine-structure constant $\alpha$ and weak electromagnetic (EM) fields. However, in the regime of strong fields, often called strong-field QED (SFQED), the perturbative methods break down and non-perturbative QED predicts many phenomena that are yet to be confirmed experimentally (for a recent review see~\cite{Review}). This motivates both particle physicists and laser physicists around the world to conduct ground-breaking experiments using high-power laser techniques \cite{E144_1999, AstraGemini_2018, ELINP_2017, E320_2019}. This synergy between the two communities, particle and laser physicists resulted in the proposed new experiment LUXE ({\bf L}aser {\bf U}nd {\bf X}fel {\bf E}xperiment) at DESY using a high-energy and high-quality intense electron beam from the European XFEL (Eu.XFEL) colliding with a high-power laser beam. The LUXE collaboration is a growing international collaboration of, presently, about 90 members. Details about the LUXE Conceptual Design Report (CDR) can be found in~\cite{CDR}. 

The flow of this report is as follows. First the definition of some parameters used in the LUXE experiment will be given. This will be followed by some specifications of the laser and the Eu.XFEL electron beam. For the understanding of the main goals of LUXE, the physics of two non-linear processes will be presented together with the challenges of measuring predictions of non-perturbative QED in a region never reached before. After a general description of the detectors planned to be used for these measurements, the report will end with conclusions and outlook.

\subsection{Meaning of non-perturbative QED}
Before going into the details of the LUXE experiment, I would like to point out that non-perturbative calculations in QED have a different meaning than those in non-perturbative QCD. 
In the case of QCD, the non-perturbative region is a kinematic regime where the theoretical calculations that need to include higher-order terms do not converge, like when the strong coupling $\alpha_S$ is too large. The calculations break down and we have non-perurbativity in $\alpha_S$.

In the SFQED regime, we do not probe non-perturbativity
due to the running of $\alpha$ QED;
we probe non-perturbativity in the coupling of matter to the laser field, which is characterised 
by a dimensionless coupling $\xi$ (to be fully defined below) which can greatly exceed unity. As such, processes and diagrams at a fixed order in $\alpha$ depend on all powers of $\xi$; effective resummation in $\xi$ at fixed $\alpha$ is however analytically possible through the use of particular models of the laser field geometry. More details follow when describing the processes.

\subsection{The Schwinger limit }
In the QED vacuum, fluctuations occur causing virtual electron-positron pairs to be constantly created and annihilated. When a pair is created, the electron and the positron initially separate, then reunite into annihilation. If during the fluctuation time a high enough external energy source $E > 2m_e$ ($m_e$ - electron mass) is supplied, for example by an external strong electric field ${\cal E}_{crit}$, the distance between the pair will be larger than the electron Compton wavelength $\hbar/{m_e c}$ and a real electron-positron pair will be produced out of the vacuum. 
Virtual photons spontaneously convert into real electron-positron pairs at a critical field, quoted as the Schwinger limit \cite{Sauter_1931, EulerHeisenberg_1936, Schwinger_1951}, 
\begin{equation}
{\cal E}_{crit} = \frac{m_e^2 c^3}{\hbar e}
	= 1.32 \times {10}^{18} {V/m}.
\end{equation}
The fluctuating vacuum is stimulated by a high electric field ${\cal E} \geq {\cal E}_{crit} $
to produce real $e^+e^-$ pairs~(see eg \cite{acatrinei}).

The Schwinger limit has not yet been reached experimentally. In fact, its value is orders of magnitude higher than the field of the most powerful lasers available today. However, in the rest frame of a high-energy probe charge, the EM field strength ${\cal E}$ is boosted by the Lorentz factor $\gamma$, to ${\cal E}^*$ = $\gamma {\cal E}(1+\cos\theta$), where $\theta$ is the collision angle between the electron and the laser beams. With a 16.5 GeV electron beam and $\theta$=17.2$^{\circ}$, resulting in $\gamma$ = 3.3 $\times$ 10$^4$, it is possible to reach the Schwinger limit in the electron rest frame with laser power of the order of 40 - 350 TW, as planned for LUXE.

\subsection{Goals of the LUXE Experiment}
The LUXE experiment, has the goal to precisely measure the physics in the non-perturbative QED regime, including (1) the interactions of laser photons with GeV electrons and photons around the critical field, (2)  in the transition regime, as well as (3) the possible creation of particles beyond the Standard Model (such as axion-like-particles) in an intense photon beam. The experiment will be conducted by observing the collision between a multi-TW optical laser and a high-quality GeV electron or a converted photon beam from the Eu.XFEL. Such collisions will allow LUXE to study the non-perturbative physics near or even above the Schwinger limit in the rest frame of the produced pair. LUXE will be one of the first experiments to reach the uncharted regime of QED with its first data to be taken in 2025/6. So far it is the only planned experiment to study the interactions of high-energy real photons with laser photons.

\section{Definition of some kinematic variables}

\begin{figure}[h!]  
   \centering
   \includegraphics[width=6.5cm]{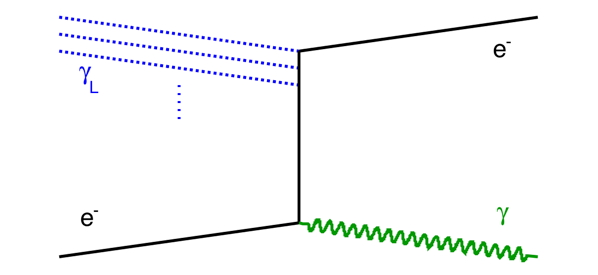} 
   \caption{A Feynman diagram of a Compton scattering process resulting from electron - laser collision.}
   \label{fig:Compton}
\end{figure}

In high energy particle physics we are used to look at collisions between two single particles. The kinematic variables are defined accordingly. However when a beam particle interacts with an intense laser beam, interactions with multi-photons at a given time have also to be considered. An example of a Feynman diagram describing the Compton scattering process of an electron beam with laser photons is shown in Fig.~\ref{fig:Compton}.
The figure shows a single electron interacting with $n$ optical photons $\gamma_L$
from the laser and converts them into a single, high-energy Compton photon,
\begin{equation}
e + n\gamma_L \to e + \gamma.
\label{eq:Compton}
\end{equation}
The three lines labelled $\gamma_L$ represent $n$ optical laser photons, where $n$ may be one (linear Compton) or a larger number, $n \geq 2$ (non-linear Compton).


One introduces dimensionless kinematic variables, also called 'parameters', which help to characterise whether a process can be described by perturbative or by non-perturbative QED. 

\subsection{Charge field coupling $\xi$ - 
the classical nonlinearity parameter }
What separates strong-field QED phenomena from regular QED is the dependency on the dimensionless charge-field coupling. It can be defined as the work done by the EM field ${\cal E_{\rm{L}}}$ over an electron Compton wavelength in units of the EM field photon energy,
\begin{equation}
\xi = |e|
{\cal E_{\rm{L}}}\frac{\lambdabar_e}{\hbar \omega_L} = |e| \frac {\cal E_{\rm{L}}}
{m_e c \omega_L},
\end{equation}
where $e$ is the charge of an electron, $\lambdabar_e$ = $\hbar/(m_e c)$ the reduced Compton wavelength of an electron and $\hbar \omega_L$ the
energy of a laser photon. The parameter is classical because it is independent of $\hbar$.

The parameter $\xi$ gives a measure of the number of laser photons interacting with the electron at a given time.  The value of $\xi^2$ is a measure of the photon density of the laser beam. It can also be expressed as
\begin{equation}
    \xi = \frac{m_e c^2}{\hbar \omega_L}\frac{\cal E_\text{L}}{{\cal E}_\text{crit}}.
\end{equation}
Whenever $\xi \ll 1$, the 
perturbative methods can be used for calculating the probability of the process. 

\subsection{ The quantum nonlinearity parameter - $\chi$}
How much QED deviates from the classical limit is in part quantified by the quantum nonlinearity parameter $\chi$ which, for an incident electron,
can be written as 
\begin{equation}
\chi = |e|{\cal E_\text{L}}\frac{\lambdabar_e}{m_e c^2}(1 + \cos{\theta}). 
\end{equation}
It is the work done over the reduced electron Compton wavelength $\lambdabar_e$ in units of $m_e c^2$. 
It can also be written as
\begin{equation}
\chi = \gamma \frac{\cal E_\text{L}}{{\cal E}_\text{crit}}(1 + \cos{\theta}),
\end{equation}
which is the ratio of the laser EM field to the Schwinger limit,
in the rest frame of the electron. The factor $\gamma$ is the relativistic Lorentz factor. The parameter $\chi$ also quantifies the recoil of the electron in the interaction.
In kinematic regions where $\chi \geq 1$, quantum effects are significant.


\subsection{ The energy parameter - $\eta$}

The center-of-mass energy squared (usually denoted by $s$ in particle physics) in units of the pair rest energy, $2m_e c^2$, is the dimensionless energy parameter $\eta$. For the case where a probe electron of energy $E_e$ collides with a laser field of energy $E_L$ at a crossing angle $\theta$, the energy parameter is
\begin{equation}
\eta = \frac{E_e E_L}{m_e^2 c^4} (1 + \cos{\theta}).
\end{equation}


The three parameters $\xi$, $\chi$ and $\eta$ are related as follows:
\begin{equation}
    \chi = \eta \xi.
\end{equation}

\section{The LUXE Experiment}

Now that the dimensionless parameters have been defined, we would like to present some details about the specifications of the LUXE experiment.
As mentioned, an intense  electron beam of 
16.5~GeV from the Eu.XFEL, located in Schenefeld, Germany, will interact with a 40~TW high-power laser, 
 in the initial state of the experiment (Phase-0) and a 350~TW laser in the second stage (Phase-1) to study SFQED processes.

\subsection{Laser}
The high power chirped pulse amplification (CPA) laser for the LUXE Phase-0 is the 800 nm (1.55 eV) Ti:Sapphire ``JETI-40'' with the power of 40~TW and a pulse duration of 30~fs. In 
Phase-1, the laser system will be upgraded to a commercial 350~TW one. With a 3 $\mu$ m focal spot, these lasers can reach peak $\xi$ values of 7.9 and 23.6, respectively. The laser will operate at a 1 Hz repetition rate. With a high precision laser diagnostic system, the laser output is expected to have less than 5\% peak uncertainty and 1\% shot-to-shot intensity uncertainty.

\subsection{Electron beam}
The Eu.XFEL electron beam used by LUXE is the result of electrons that are injected at DESY, Hamburg and accelerated through a 1.9~km linear accelerator. 
Electron-bunch trains with up to 2700 bunches are injected 10 times per second into the linac and accelerated to 16.5 GeV; one bunch per train, containing typically $1.5\times 10^9$ electrons, is extracted and guided into the LUXE tunnel.
Since the laser is a 1Hz beam, 9 of the remaining non-interacting bunches will serve to study the background {\it in-situ}. The electron beam can directly interact with the laser, or be converted to GeV photons via bremsstrahlung. The collision angle is set at 17.2$^\circ$ . After the boost, $\chi$ can reach a value of 1.5 in Phase-0 and 4.5 in Phase-1.
\begin{figure}[h!] 
   \centering
   \includegraphics[width=10cm]{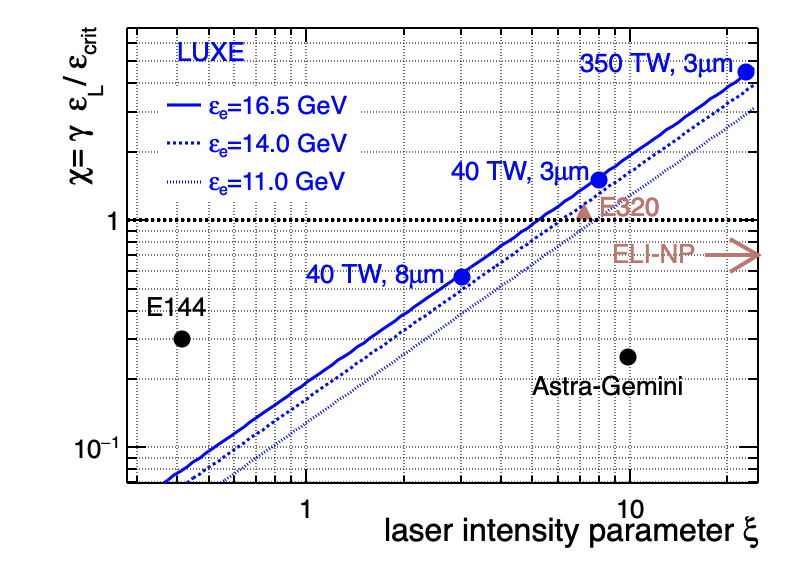} 
   \caption{Quantum parameter $\chi$ as a function of the intensity parameter $\xi$ for LUXE and a selection of experiments and facilities \cite{CDR}.}
   \label{fig:chi_xi}
\end{figure}
Figure~\ref{fig:chi_xi} shows the proposed LUXE experiment in the kinematic plane of  $\chi$ - $\xi$, together with the landmark SLAC E144 experiment from the mid 1990s~\cite{E144_1999}, and additional future planned experiments.
The E144 experiment fell short of the Schwinger limit, 
measured processes
in the non-linear, multi-photon regime but remained perturbative, $\xi < 1$. 
The E320 experiment~\cite{E320_2019} is planning to reach the $\xi \geq 1$ limit in the near future. LUXE will be able to study the transition region and reach the non-linear region. It will also be the only such experiment that will collide high energy photons with a laser beam ($\gamma$-laser mode).

\section{Non-linear processes}
\subsection{Non-linear Compton Scattering}
\begin{figure}[h!]  
   \centering
    \hspace*{0.5cm}
   \includegraphics[width=4cm]{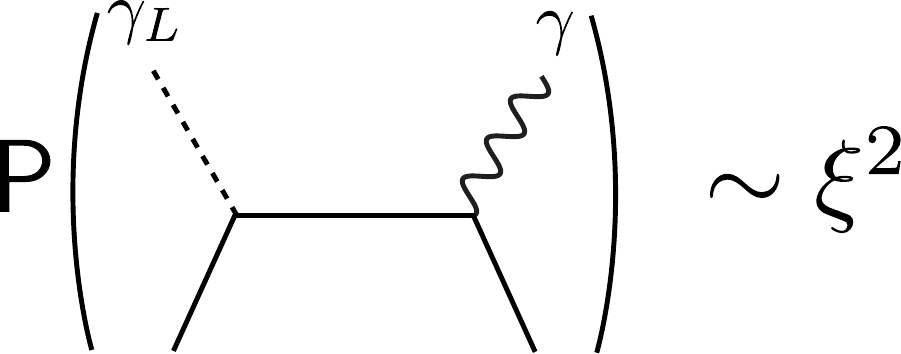} 
   \hspace*{0.5cm}
   \includegraphics[width=4cm]{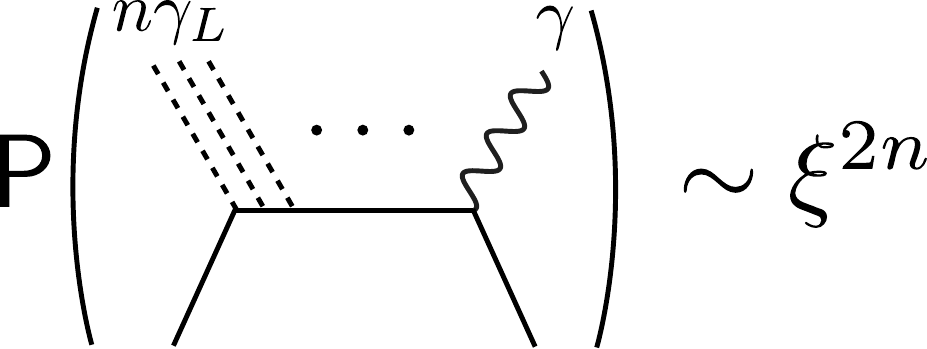}
    \hspace*{0.5cm}
   \includegraphics[width=4cm]{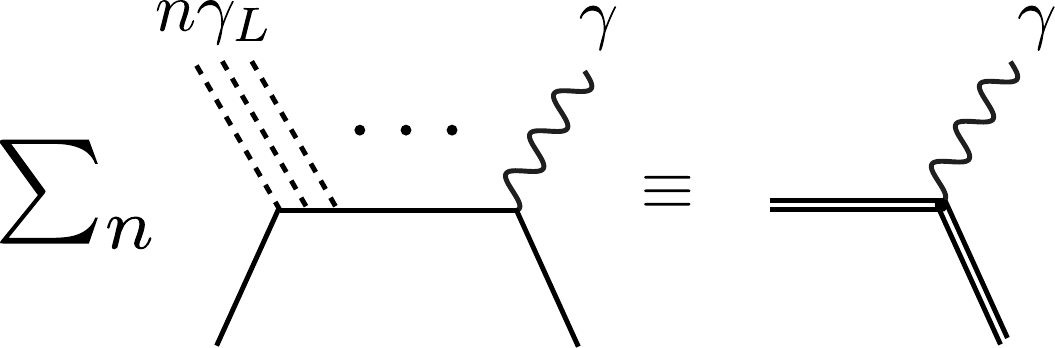}
   \caption{A diagram describing the probabilities of a Compton scattering process resulting from an electron  colliding with (left) one laser photon $\gamma_L$, (linear Compton), (center) n laser photons n$\gamma_L$, (non-linear Compton). The right diagram presents the sum of the Compton scattering at large densities ($\xi \sim 1)$ resulting in a non-linear, non-perturbative Compton process.}
   \label{fig:C-prob}
\end{figure}

The diagram shown in Fig.~\ref{fig:Compton} describes a Compton scattering process. When the charge-field coupling is small, meaning low laser photon density, and the electron collides with just one laser photon $\gamma_L$ ($n$=1 in Eq.~(\ref{eq:Compton})), the amplitude is linear in $\xi$ (linear Compton scattering), and the probability $P$ to produce one Compton photon is proportional to $\xi^2$ (Fig.~\ref{fig:C-prob} (left)). Still at low laser photon densities ($\xi < 1$), the electron can collide with $n$ laser photons to produce a single high energy Compton photon. In this case the probability is proportional to $\xi^{2n}$ and the Compton process is non-linear (Fig.~\ref{fig:C-prob} (center)). At low densities it is sufficient to consider this number of laser photons and the procedure is still perturbative.
However, at high laser-photon density ($\xi \geq 1$), there are no more leading-order processes and
one has to sum up all higher-order contributions (Fig.~\ref{fig:C-prob} (right)) which leads into the non-perturbative region. The resulting expression can be interpreted, roughly, as if the electron has an effective mass which increases with $\xi^2$,
\begin{equation}
    m_e(\text{eff}) = m_e \sqrt{1 + \xi^2}.
\end{equation}
The $\xi$ dependence on the electron effective mass causes a shift of the 'Compton edge'~\cite{Compton-edge} in the electron and photon energy spectra.

\subsubsection{Compton edge}
\begin{figure}[h!]
   \centering
   \includegraphics[width=10cm]{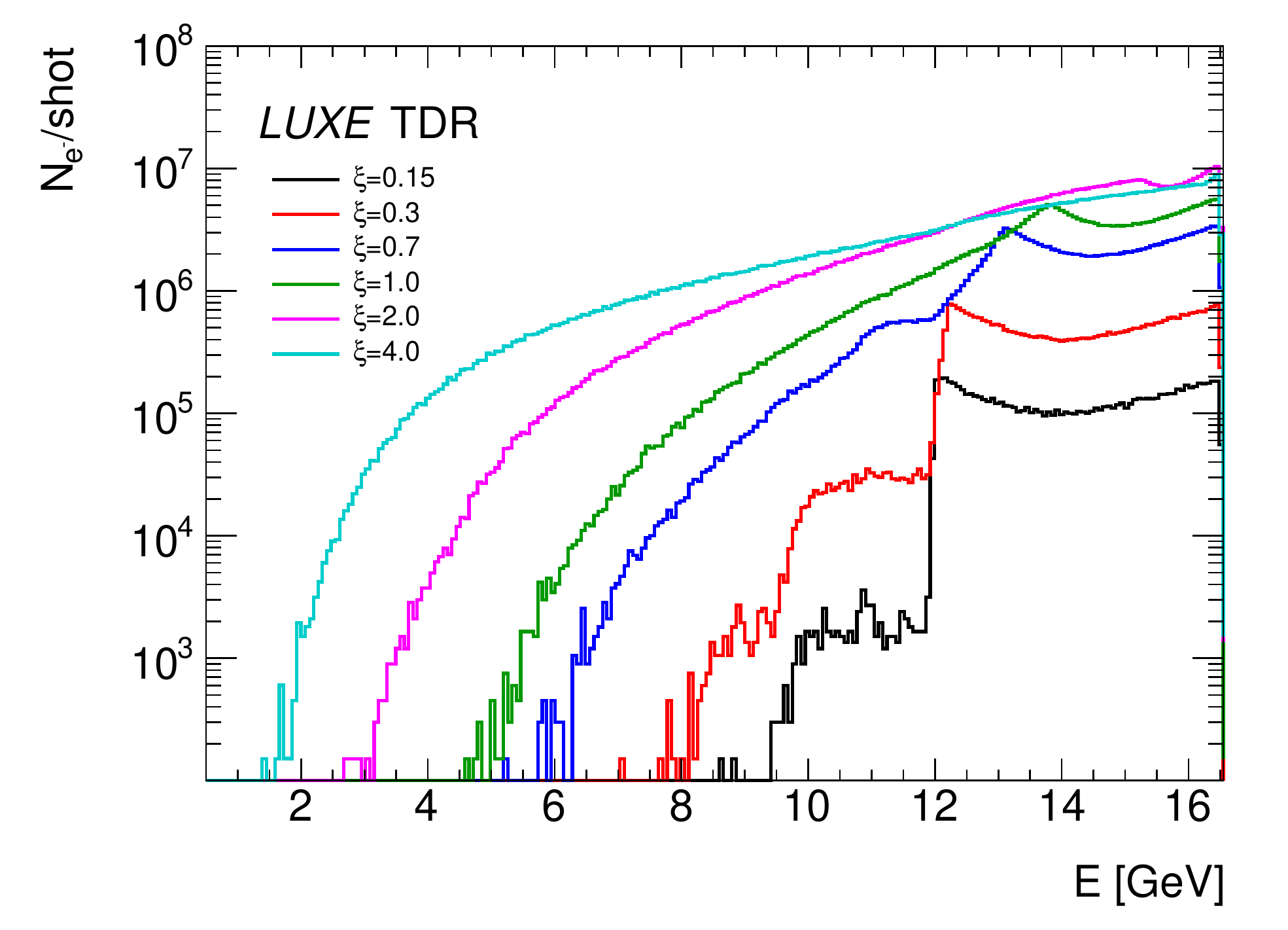} 
   \includegraphics[width=10cm]{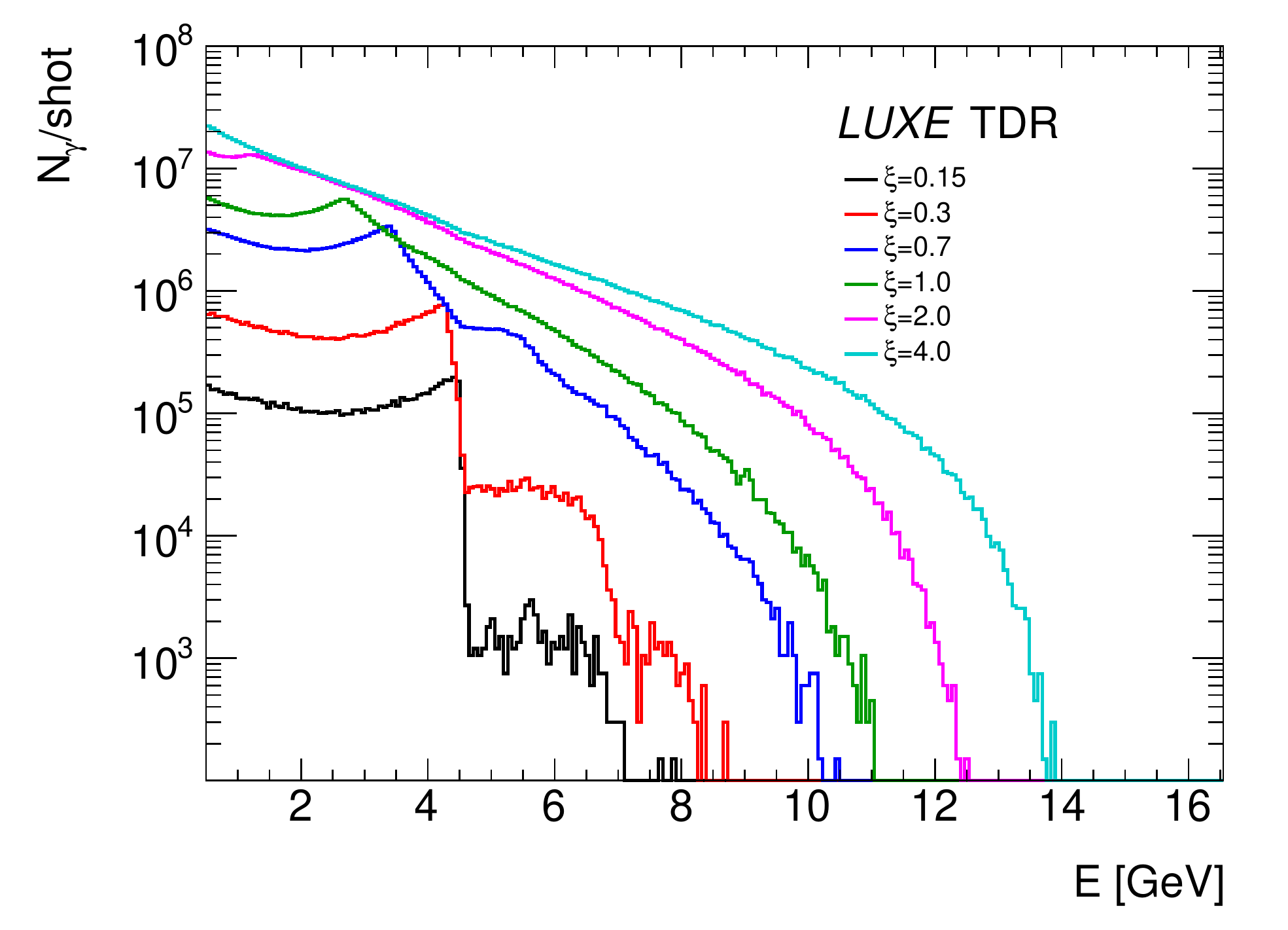} 
   \caption{Number of electrons (upper) and photons (lower) per bunch crossing for the Compton process as function of the respective energy, based on calculations for a selection of $\xi$ values.}
   \label{fig:C-shift}
\end{figure}

The electron (photon) energy spectrum for the Compton process is shown on the upper part (lower part) of Fig.~\ref{fig:C-shift}, calculated for different $\xi$ values. The energy spectra were produced using the open-source code {\sc{Ptarmigan}}~\cite{Ptarmigan}. The locally monochromatic approximation (LMA) has been developed for the use in simulating the interaction  physics~\cite{BW-Ben-Tom,Tom-AJ-Ben}, with limitations to the LMA being explored in~\cite{Ben,Tang-Ben}

For $\xi = 0.15$ one sees the drop after the Compton edge at 12~GeV from the interaction of the beam electron with one laser photon, followed by two orders of magnitude less electrons at an edge of about 10~GeV from the interaction with two laser photons. Contributions of electrons from scattering with more than two photons produces negligible amounts of electrons. For $\xi =  0.3$ one can clearly see the edges after
absorbing a net number of photons, $n = 1, 2$ \& $3$. Once we get to higher values of $\xi$, the higher-order terms contribution can not be neglected and one cannot see anymore clear Compton edges. However, the figure shows clearly how the energy value of the first Compton edge moves to higher electron energies. Accordingly it moves to lower energies for the Compton photon energies. 

For the electron energy, the shift of the first edge is predicted to move as follows:
\begin{equation}
    E_{edge}^e(\xi) = E_e \frac{1 + \xi^2}{2\eta + 1 + \xi^2}.
\end{equation}
For the Compton photon energy shift, $E_{edge}^\gamma(\xi)$, the numerator $1 + \xi^2$ is replaced by $2 \eta$.

\begin{figure}[h!]
   \centering
   \includegraphics[width=10cm]{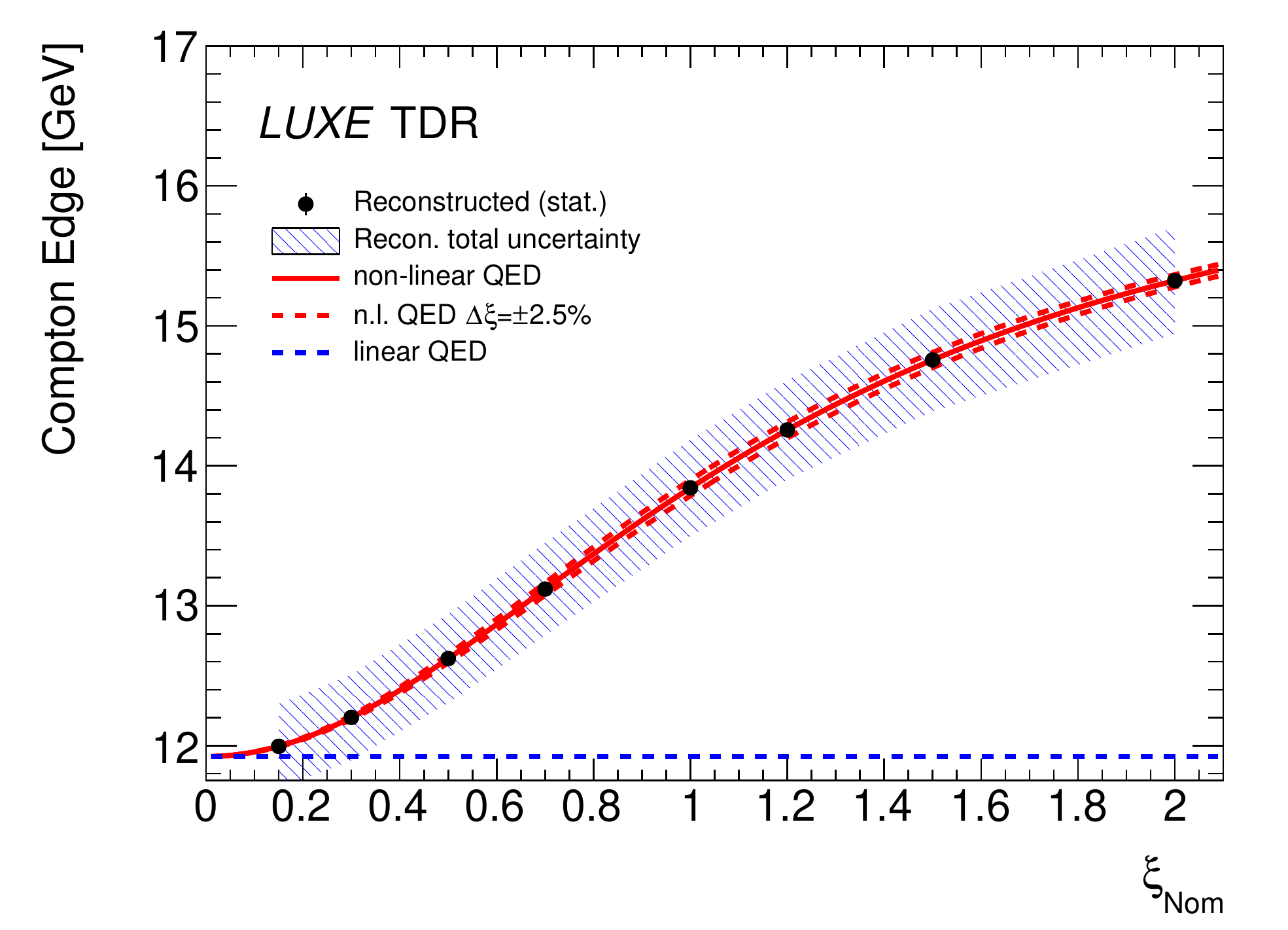} 
   \caption{Expected reconstructed electron Compton edge positions as a function of $\xi$, compared to linear and non-linear QED predictions. The band indicates a 2.5\% uncertainty on the energy scale. Statistical errors based on one hour of data-taking are shown but negligible.}
   \label{fig:C-shift-pred}
\end{figure}
The expected shift from linear and non-linear Compton scattering as a function of $\xi$ is shown in Fig.~\ref{fig:C-shift-pred}. Also shown are the reconstructed values of the Compton edge positions in the planned detector (to be discussed later), assuming a 2.5\% uncertainty on the energy scale.
One can see a strong deviation of the non-linear expectations from the linear QED result, the latter being $\xi$ independent. This measurement would be a clear signal of reaching the non-perturbative SFQED regime.

\subsection{Non-linear Breit-Wheeler}
\begin{figure}[h]
   \centering
   \includegraphics[width=6cm]{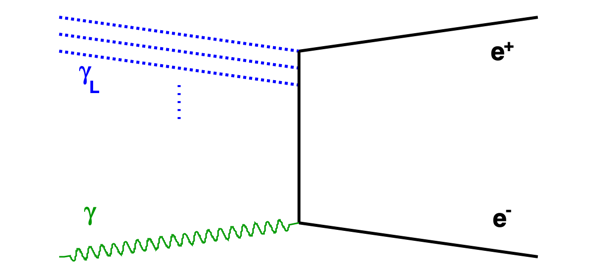} 
   \caption{Feynman diagrams of the Breit--Wheeler process. }
   \label{fig:BW}
\end{figure}

The non-linear Breit-Wheeler process~\cite{BW,NR} is described by the Feynman diagram in Fig.~\ref{fig:BW}.
In the non-linear Breit-Wheeler process a real photon (produced either in non-linear Compton
scattering in the LUXE e-laser mode, 
or via Bremsstrahlung in a target in the $\gamma$-laser mode), absorbs multiple laser photons and produces a physical electron-positron pair. 
Like in the Compton scattering diagram, the three lines labelled $\gamma_L$ represent $n$ optical laser photons.
The reaction can be written as
\begin{equation}
\gamma + n\gamma_L \to e^+ + e^-,
\label{eq:BW}
\end{equation}
where $n$ is the total number of photons absorbed.

\begin{figure}[h!]  
   \centering
   \includegraphics[width=4cm]{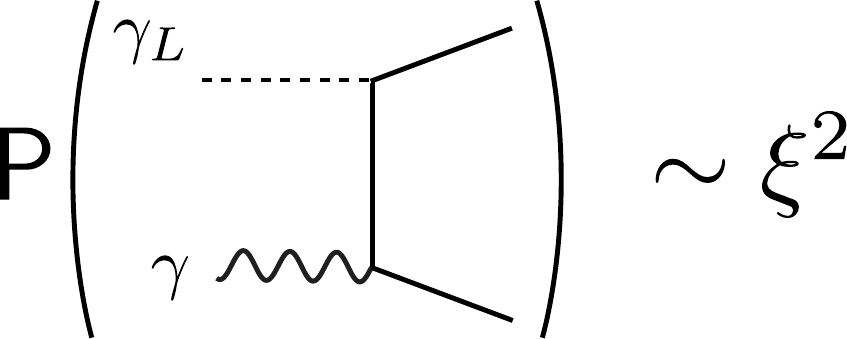} 
   \hspace*{0.5cm}
   \includegraphics[width=4.5cm]{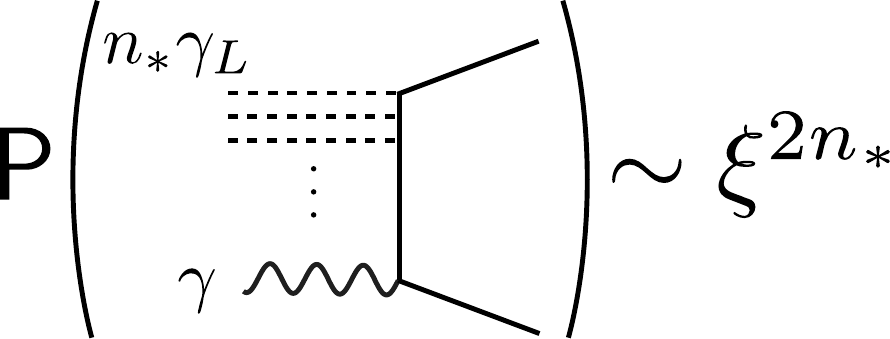}
    \hspace*{0.5cm}
   \includegraphics[width=5cm]{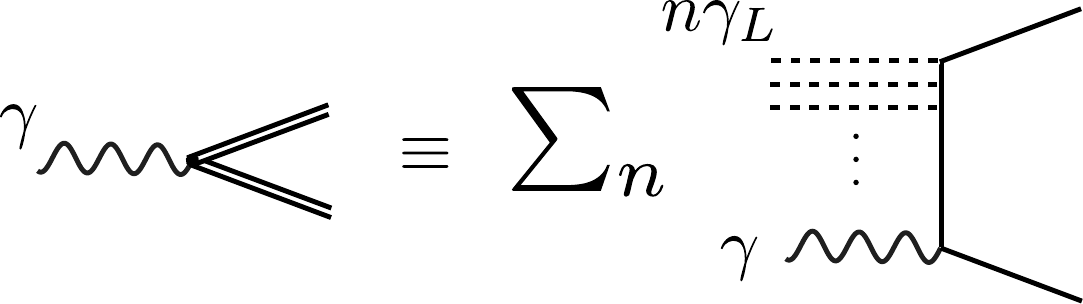}
   \caption{A diagram describing the probabilities of a Breit-Wheeler process resulting from an photon  colliding with (left) one laser photon $\gamma_L$, (linear Breit-Wheeler), (center) $n_*$ laser photons $n_* \gamma_L$, (non-linear Breit-Wheeler). The right diagram presents the sum of all order contributions to the Breit-Wheeler process at large densities ($\xi \sim 1)$ resulting in a non-linear, non-perturbative Breit-Wheeler process.}
   \label{fig:BW-prob}
\end{figure}

Contrary to  the Compton scattering case, this process can occur only if the center-of-mass energy is large enough to produce an electron-positron pair. If this energy threshold can be reached by interacting with just one laser photon the process is the linear Breit-Wheeler and the probability for it to occur is proportional to $\xi^2$ (left diagram in Fig.~\ref{fig:BW-prob}). Still at low laser intensity, the mass threshold can be reached only by the photon interaction with many laser photons, $n_*$ of them. The process is then a multi-photon non-linear process and the probability for it to happen is proportional to $\xi^{2n_*}$ (center diagram in Fig.~\ref{fig:BW-prob}). When the field strength gets large ($\xi \geq 1$), there is no more a leading-order term and all contributions have to be summed up (right diagram in Fig.~\ref{fig:BW-prob}). The Breit-Wheeler process becomes non-linear and non-perturbative (see eg~\cite{BW-Ben-Tom,Tom-AJ-Ben,Hartin:2018sha}). 

\begin{figure}[h!]
   \centering
   \includegraphics[width=8.5cm]{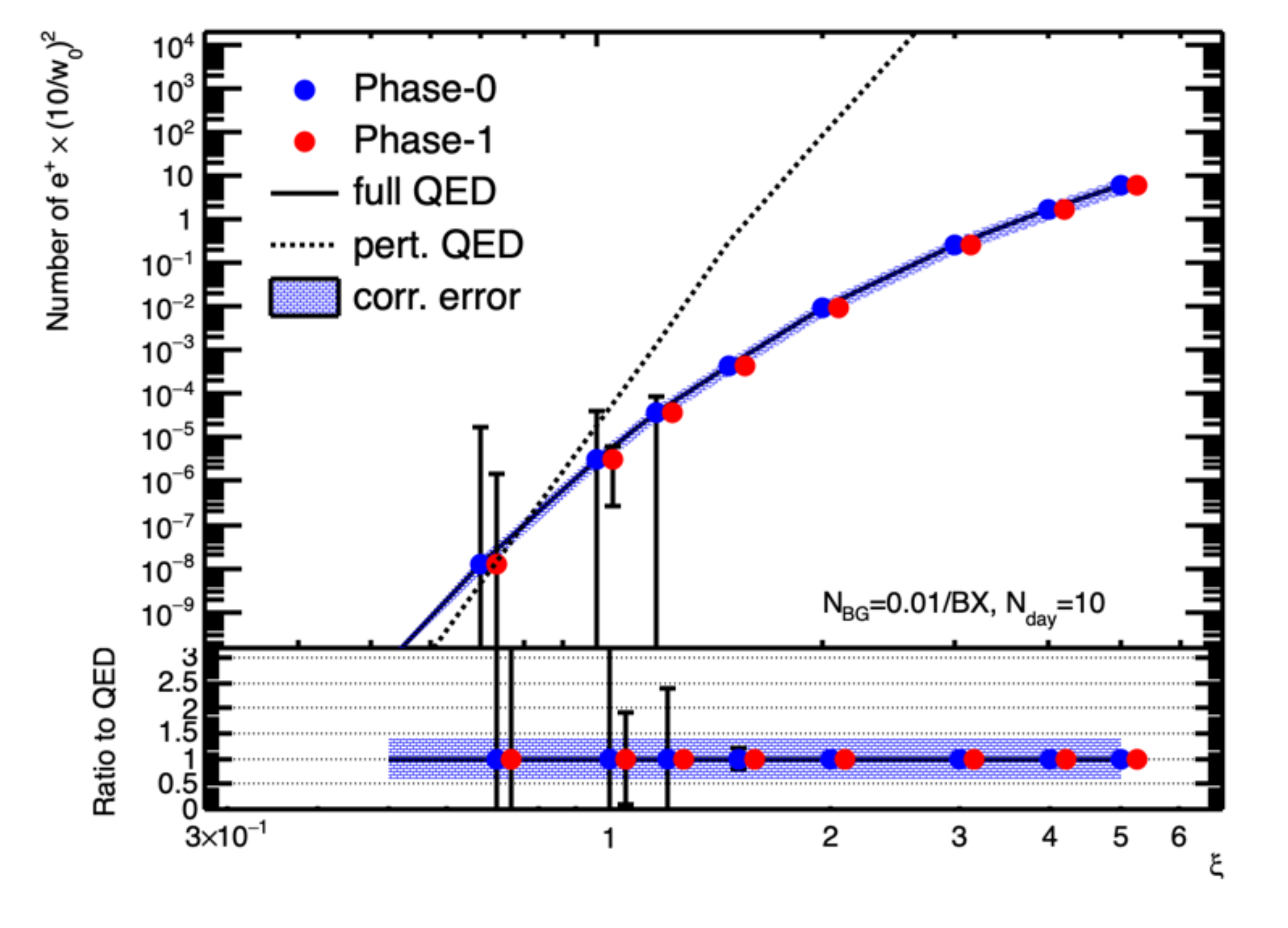}
   \caption{ Expected rate of positrons created via the Breit-Wheeler pair production per bunch crossing as a function of laser intensity in $\gamma$-laser collisions, assuming 10 days of data-taking and a background rate of 0.01 events/bunch-crossing. The solid black line are predictions of an all-order full QED calculations, while the dotted 
   line is that of perturbative QED. The blue band is the impact of a 2.5\% uncertainty on the measured value of $\xi$, resulting in an up to 40\% uncertainty on the positron rate. The lower-part of the plots shows the ratio of the reconstructed rate to the non-perturbative QED predictions for the two phases of the experiment.}
   \label{fig:BW-pos-pred}
\end{figure}

The non-perturbative calculations which sum up all-order contributions to the process, predict a slower increase of the rate of positron production than the perturbative one, as the laser intensity increases. Figure~\ref{fig:BW-pos-pred} shows the expected rate of positrons created via the Breit-Wheeler pair production per bunch crossing as a function of laser intensity in $\gamma$-laser collisions, assuming 10 days of data-taking and a background rate of 
0.01~events/bunch-crossing.  
 One observes a clear deviation of the reconstructed positron rate from that predicted by perturbative QED for $\xi$ values  already reached in Phase-0 of LUXE. 

\section{Detectors}
\subsection{Challenges}
The key plots of the two non-linear processes discussed above, Compton scattering and Breit-Wheeler pair production, that will indicate that LUXE reached the non-perturbative QED regime, were shown in Fig.~\ref{fig:C-shift-pred} and Fig.~\ref{fig:BW-pos-pred}. 

\begin{figure}[h!]
   \centering
   \includegraphics[width=9.5cm]{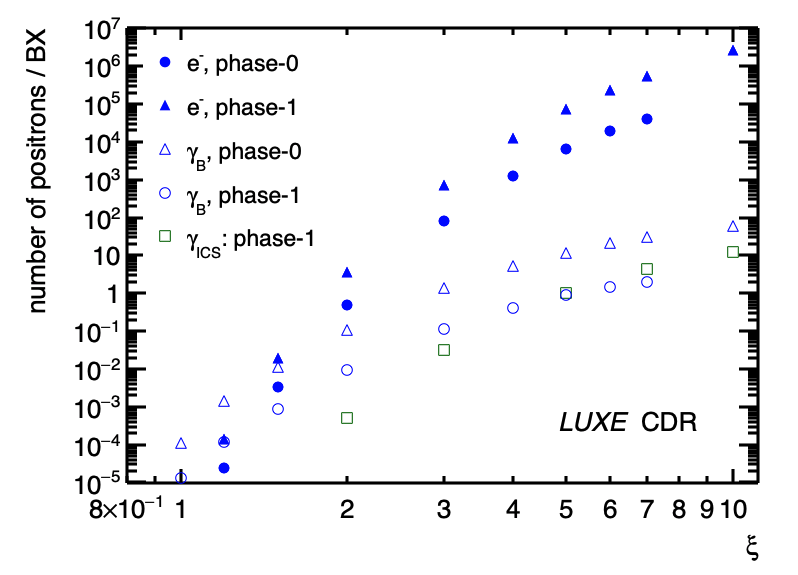} 
   \includegraphics[width=9.5cm]{SpectrumCompareXi-TomMC-ph.pdf} 
   \caption{Upper-part: number of positrons per bunch crossing produced in e/$\gamma$-laser setups as a function of $\xi$, for Phase-0 and Phase-1 of LUXE. ($\gamma_{\text{B}}$ - bremsstrahlung photons, $\gamma_{\text{ICS}}$ - inverse Compton scattering photons).
   Lower-part: Number of photons per bunch crossing for the Compton process as a function of the photon energy, based on calculations for a selection of $\xi$  values. }
   \label{fig:challenges}
\end{figure}
The challenges of LUXE are to be able to reconstruct with high precision the rates of the positrons, electrons and photons and to have a good enough energy reconstruction for measuring the Compton edge to observe its shift as a function of the laser intensity. The large number of expected positrons and photons can been seen in Fig.~\ref{fig:challenges}. While the number of expected positrons per bunch-crossing at high intensities for the $\gamma$-laser case reaches $\sim$ 100 of positrons, for the $e$-laser it may be as high as $10^6$ positrons (Fig.~\ref{fig:challenges} upper-part). Also the number of photons per bunch-crossing can reach at high intensities $\sim 10^6$ (Fig.~\ref{fig:challenges} lower-part). Theses high fluxes of particles need detectors that can meet these challenges.

\subsection{Detector components}
\begin{figure}[h!]
   \centering
   \includegraphics[width=7cm]{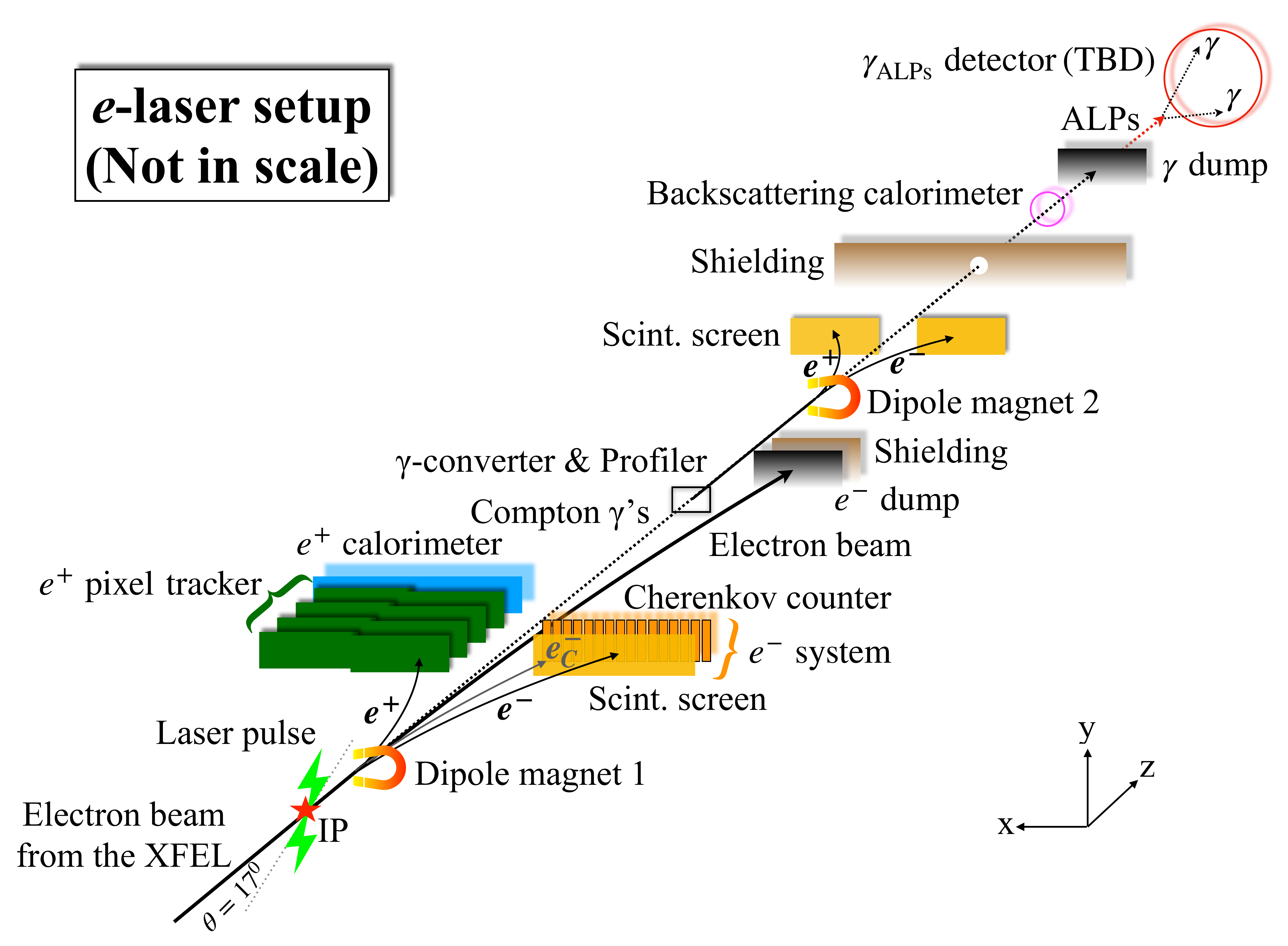}
   \includegraphics[width=7cm]{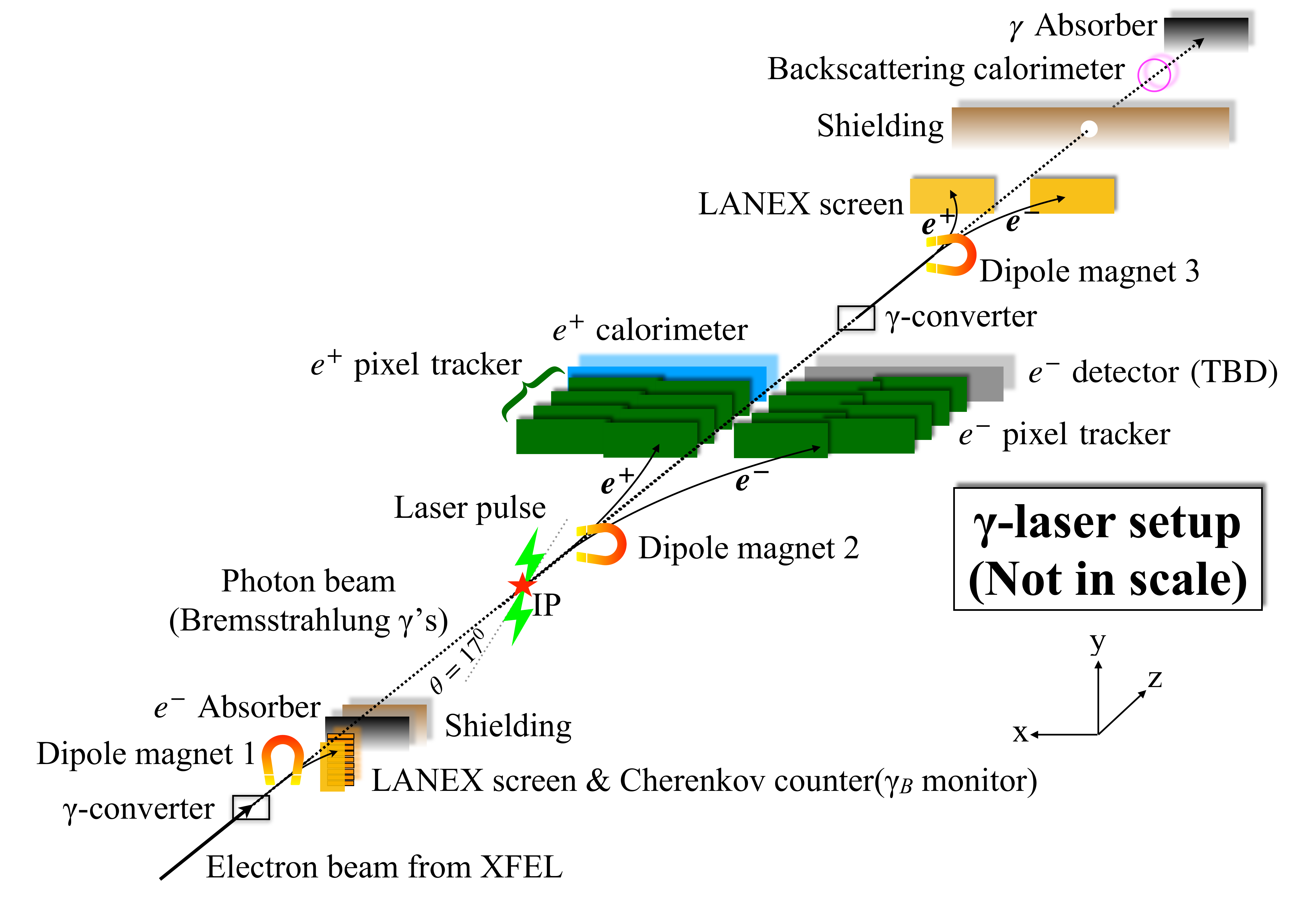} 
   \caption{Schematic LUXE experimental layouts for the electron-laser (left) and $\gamma$-laser (right) set-ups.}
   \label{fig:det-setup}
\end{figure}

\begin{figure}[h!]
   \centering
   \includegraphics[width=4.5cm]{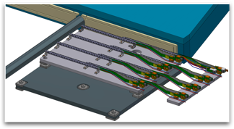} 
   \includegraphics[width=4.5cm]{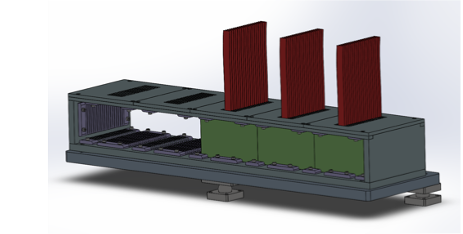} 
    \includegraphics[width=4.5cm]{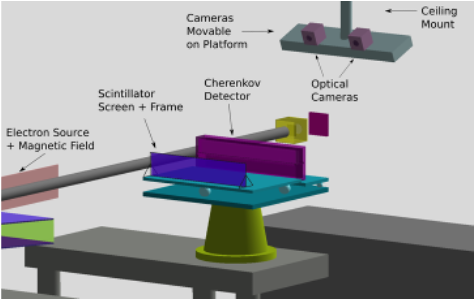} 
     \includegraphics[width=4.5cm]{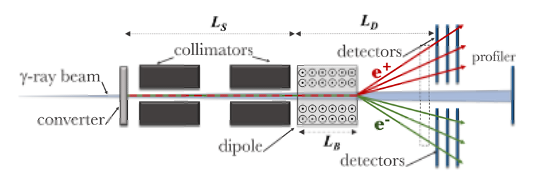} 
      \includegraphics[width=4.5cm]{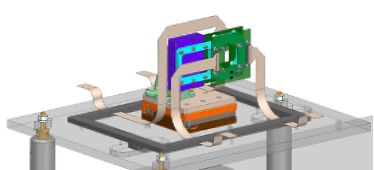} 
       \includegraphics[width=4.5cm]{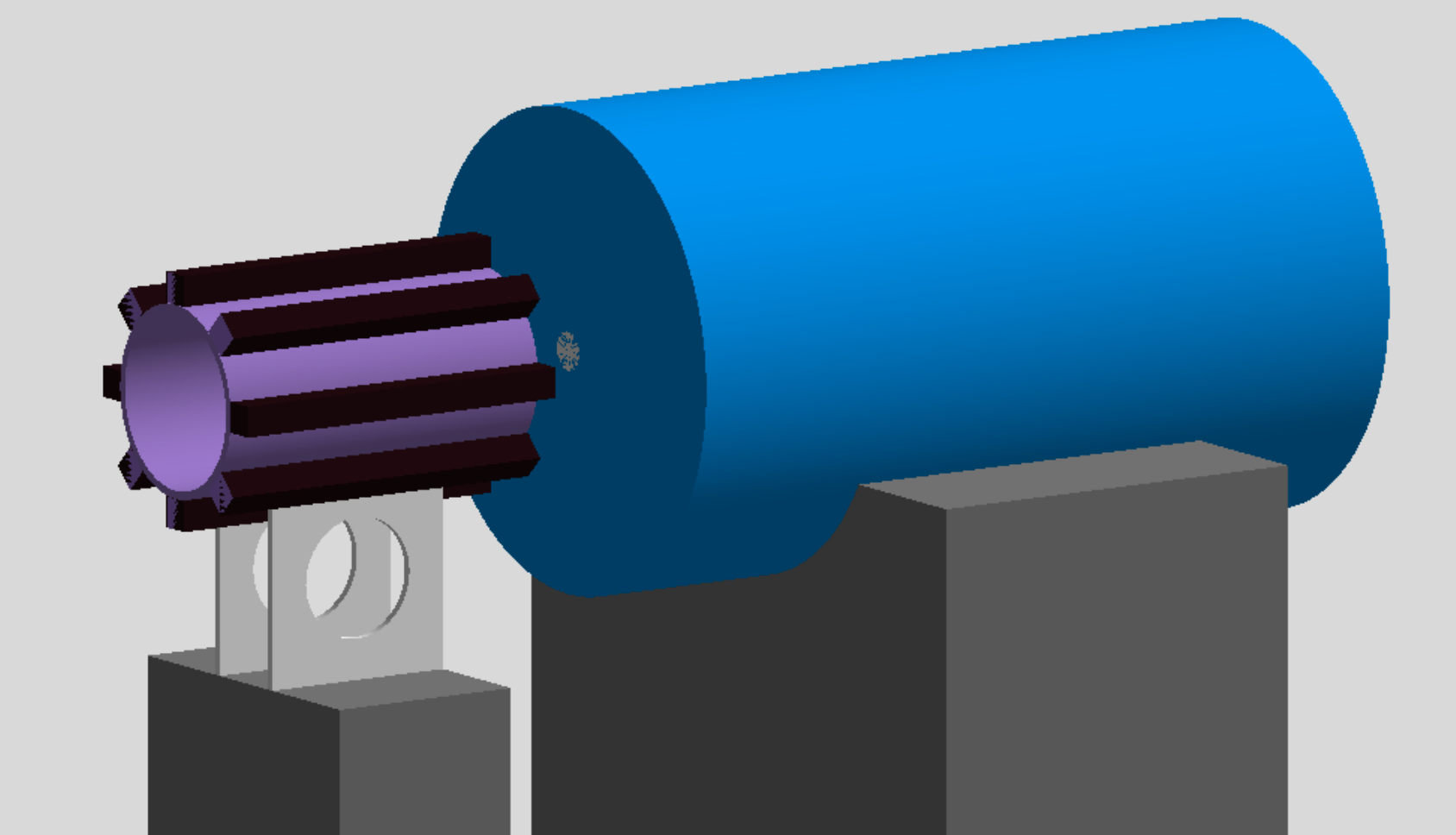} 
   \caption{Some LUXE detector components. From top left, in clock-wise direction: Tracker, Calorimeter, Cherenkov with scintillator screens, Gamma ray spectrometer, Gamma beam profiler and Gamma flux monitor. }
   \label{fig:det-components}
\end{figure}

A schematic layout of the detector components is presented in Fig.~\ref{fig:det-setup}. The left-hand side of the figure shows the setup for the electron-laser mode while the right-hand side, for the $\gamma$-laser one.
A collection of different detectors and magnets are needed for measuring the expected high flux of particles expected in the SFQED regime. These include Tracker, Calorimeter, Cherenkov, Scintillator screens, Gamma-Ray Spectrometer, Gamma-Ray Profiler, Gamma-flux Monitor. Details of all these can be found in~\cite{CDR}. Figure~\ref{fig:det-components} shows schematics of these components.
A technical design report (TDR) of each component has been written and a complete TDR of LUXE is to be published soon~\cite{TDR}.

\section{Bonus: searching for BSM with LUXE}
\begin{figure}[h!]
   \centering
   \includegraphics[width=8cm]{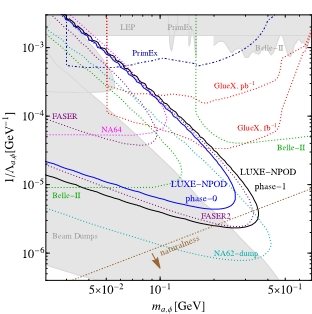}
   \caption{ The projected reach of the LUXE-NPOD proposal in the effective coupling \newline ($\frac{1}{\Lambda_X}$ ) vs the mass ($m_X$) parameter space. The phase 0 (1) result is shown in blue (black). The currently existing bounds are shown in grey region.}
   \label{fig:alps}
\end{figure}

In addition to probing strong-field QED physics processes, the enormous rate of photons produced in LUXE (mainly through non-linear Compton scattering in the $e$-laser setup) can be used to search for physics beyond the Standard Model (BSM). 
One possible scenario is axion-like particles (ALPs) being created via Primakoff production in the LUXE photon beam dump. By placing a calorimeter at a fixed distance behind the photon dump and ensuring a low background in this detection area, decays of the ALPs to two photons can be probed. With a 1 $\times$ 1 m$^2$ detector, a sensitivity can be achieved that exceeds that of other existing experiments searching for ALPs in the 100~MeV range (See Fig.~\ref{fig:alps}). 
This aspect of LUXE is called LUXE-NPOD (LUXE New Physics search with Optical Dump) and described in
more detail in~\cite{NPOD}.

\section{Conclusions and Outlook}
The LUXE experiment provides an
exciting opportunity to explore QED in a new regime using the electron beam of the European XFEL and a high power laser.
LUXE will observe the transition from perturbative to non-perturbative QED, a region never reached before. In addition, the experiment can be used for BSM physics.

The LUXE CDR received CD0 approval by DESY.
The TDR is presently reviewed.
If approved, installation could happen in 3-4 years. Results – still in this decade!

Come and join us in a breakthrough experiment!!

\section*{Acknowledgement}
This work was partly supported by the German-Israel Foundation (GIF) and the PAZY Foundation.
I would like to thank my LUXE colleagues for their help in producing this write-up. I would like in particular to single out Ben King for his infinite patience in explaining the intricacies of the SFQED.
We thank the DESY technical staff for continuous assistance and the DESY directorate for their strong support
and the hospitality they extend to the non-DESY members of the collaboration. This work has benefited from
computing services provided by the German National Analysis Facility (NAF) and the Swedish National Infrastructure for Computing (SNIC).

{}


\begin{thebibliography}{}
\bibitem{Review} A. Fedotov et al., Advances in QED with intense background fields, arXiv:2203.00019, (2022)
\bibitem{E144_1999} C. Bamber et al., Studies of nonlinear QED in collisions of 46.6 GeV electrons with intense laser pulses, Phys. Rev. {\bf D 60} (1999) 092004
\bibitem{AstraGemini_2018} K. Poder et al., Experimental signatures of the quantum nature of radiation reaction in the field of an ultraintense laser, Phys. Rev. {\bf X 8} (2018) 031004
\bibitem{ELINP_2017} S. Ataman et al., Experiments with combined laser and gamma beams at ELI-NP, AIP Conf. Proc. {\bf 1852} (2017) 070002
\bibitem{E320_2019} S. Meuren, Probing strong-field QED at FACET-II (SLAC E-320), Talk presented at FACET-II Science Workshop 2019 
\bibitem{CDR} 
H. Abramowicz et al., Conceptual Design Report for the LUXE Experiment, Eur. Phys. J. Spec.Top. {\bf 230} (2021) 2445
\bibitem{Sauter_1931} F. Sauter, Ueber das Verhalten eines Elektrons im homogenen elektrischen Feld nach der relativistischen Theorie Diracs, Z. Phys. {\bf 69} (1931) 742

\bibitem{EulerHeisenberg_1936} W. Heisenberg and H. Euler, Folgerungen aus der Diracschen Theorie des Positrons, Z. Phys. {\bf 98} (1936) 714
\bibitem{Schwinger_1951} J. Schwinger, On gauge invariance and vacuum polarization, Phys. Rev. {\bf 82} (1951) 664
\bibitem{acatrinei} C.S. Acatrinei, QED in strong electromagnetic backgrounds – effective action methods, Rom. J. Phys. {\bf 64} (2019) 113
\bibitem{Compton-edge} C. Harvey, T. Heinzl, and A. Ilderton, Signatures of High-Intensity Compton Scattering, Phys. Rev. {\bf A 79} (2009) 063407
\bibitem{Ptarmigan} T.G. Blackburn, {\sc{Ptarmigan}}, Simulate the interaction between a high-energy particle beam and an intense laser pulse, including the classical dynamics and strong-field QED processes. https://github.com/tgblackburn/ptarmigan
\bibitem{BW-Ben-Tom} T.G. Blackburn and B. King, Higher fidelity simulations of nonlinear Breit-Wheeler pair creation in intense laser pulses, Eur. Phys. J. {\bf C 82} (2022) 1, arXiv:2108.10883, doi.org/10.1140/epjc/s10052-021-09955-3
\bibitem{Tom-AJ-Ben} T.G. Blackburn, A.J. MacLeod, and B. King, From local to nonlocal: higher fidelity simulations of photon emission in intense laser pulses, New J. Phys. {\bf 23} (2021) 085008, arXiv:2103.06673 

\bibitem{Ben} B. King, Interference effects in nonlinear Compton scattering due to pulse envelope, Phys. Rev. {\bf D 103} (2021) 036018
\bibitem{Tang-Ben} S. Tang and B. King, Pulse envelope effects in nonlinear Breit-Wheeler pair creation,
Phys. Rev. {\bf D 104} (2021) 096019
\bibitem{BW} G. Breit and J.A. Wheeler, Collision of two light quanta, Phys. Rev. {\bf 46} (1934) 1087
\bibitem{NR} A.I. Nikishov and V.I. Ritus, Quantum processes in the field of a plane electromagnetic wave and in a constant field, Sov. Phys. JEPT {\bf 19} (1964) 529; Sov. Phys. JEPT {\bf 19} (1964) 1191

\bibitem{Hartin:2018sha}
	A.~Hartin, A.~Ringwald and N.~Tapia,
	Measuring the Boiling Point of the Vacuum of Quantum Electrodynamics,
	Phys. Rev. {\bf D 99} (2019) 036008,
	doi:10.1103/PhysRevD.99.036008,
	arXiv:1807.10670 
\bibitem{TDR} H. Abramowicz et al., Technical Design Report for the LUXE Experiment, in preparation (2022)
\bibitem{NPOD} K. Bai et al., LUXE-NPOD: new physics searches with an optical dump at LUXE, arXiv:2107.13554 (2021)

\end{thebibliography}

\end{document}